\documentclass[aps,twocolumn,showpacs]{revtex4}
\usepackage{graphicx}
\usepackage{amsmath}
\usepackage{amsfonts}
\usepackage{color}
\newcommand{\tbox}[1]{\mbox{\tiny #1}}

\begin{document}



\title{Singular-value statistics of directed random graphs}


\author{J. A. M\'endez-Berm\'udez}
\address{Instituto de F\'isica, Benem\'erita Universidad Aut\'onoma de Puebla, 
Puebla 72570, Mexico \\
Escuela de F\'isica, Facultad de Ciencias, Universidad Nacional Aut\'onoma de Honduras, Honduras}

\author{R. Aguilar-S\'anchez}
\address{Facultad de Ciencias Qu\'imicas, Benem\'erita Universidad Aut\'onoma de Puebla,
Puebla 72570, Mexico}

\begin{abstract}
Singular-value statistics (SVS) has been recently presented as a random matrix theory
tool able to properly characterize non-Hermitian random matrix ensembles [PRX Quantum {\bf 4}, 040312 (2023)].
Here, we perform a numerical study of the SVS of the non-Hermitian adjacency matrices
$\mathbf{A}$ of directed random graphs, where $\mathbf{A}$ are members of diluted real 
Ginibre ensembles. 
We consider two models of directed random graphs: Erd\"os-R\'enyi graphs and random regular graphs.
Specifically, we focus on the ratio $r$ between nearest neighbor singular values 
and the minimum singular value $\lambda_{\tbox{min}}$.
We show that $\langle r \rangle$ (where $\langle \cdot \rangle$ represents ensemble average) 
can effectively characterize the transition between mostly 
isolated vertices to almost complete graphs, while the probability density function of $\lambda_{\tbox{min}}$ 
can clearly distinguish between different graph models.
\end{abstract}

\maketitle

\section{Preliminaries}

There is already a vast number of applications of Random Matrix Theory (RMT) measures and 
techniques to the study and characterization of complex graphs and networks.
The general idea of RMT is that given a matrix representing a system or process, if that system
or process is complex enough, the corresponding matrix can be substituted by an ensemble of
random matrices having the symmetries of the original matrix (see e.g.~\cite{ABF15}).
Then, luckily, such a random matrix ensemble may allow for an analytical (or a phenomenological) 
statistical approach that may describe the {\it universal} properties of the original complex system
or of the family of complex systems represented by matrices sharing the same symmetries.
Otherwise, a statistical numerical study of the random matrix ensemble may also provide useful 
information about the corresponding complex system.
In this respect, the application of RMT to complex graphs and networks is quite straightforward
through their matrix representations (see e.g.~\cite{E11}): adjacency matrix, Laplacian matrix, 
incidence matrix, etc., and variants of them.

Among the wide amount of available studies of graphs and networks from a RMT perspective we 
can mention that:
(i) the density of states of random graphs has been approached by generalizing the 
semicircle law (see e.g.~\cite{K15,B19,SJ18}),
(ii) spectral properties of several models of random graphs have been characterized 
by the use of the nearest neighbor spacing distribution, the singular value decomposition, and the 
ratio between nearest- and next-to-nearest neighbor eigenvalues, and other RMT measures 
(see e.g.~\cite{SJ18,BJ07a,BJ07b,BJ08,BJ09,MAMRP15,TFM20,MRJ22,RJ22,RJ23,AMGM18,AMRS20,MMMPS19,MMTM21,MFRM17,PRRCM20,PM23}),
(iii) eigenvector properties of several models of random graphs have been 
characterized by the use of Shannon entropies and inverse participation ratios 
(see e.g.~\cite{JSVL10,JP18,FMRM20,AMGM18,AMRS20,MMMPS19,MMTM21,MFRM17,PRRCM20,PM23,CSTV24}), while
(iv) scattering and transport properties of tight-binding random graphs have been 
studied by means of the effective non-Hermitian Hamiltonian approach 
(see e.g.~\cite{MAM13,MHM24}).

Specifically, among the models of graphs and networks which have been approached with the 
above measures and techniques we have: 
Erd\"os-R\'enyi random graphs (see e.g.~\cite{BJ07a,BJ07b,BJ08,MAMRP15,CSTV24}), 
small-world networks (see e.g.~\cite{BJ07a,BJ07b,BJ08}), 
scale free networks (see e.g.~\cite{BJ07a,BJ07b,BJ08}), 
random geometric graphs (see e.g.~\cite{AMGM18,AMRS20,PM23}), 
bipartite graphs (see e.g.~\cite{MMMPS19}), 
mutualistic graphs (see e.g.~\cite{MMTM21}), 
multilayer and multiplex networks (see e.g.~\cite{MFRM17,JP18,FMRM20,RJ22,RJ23}), etc.

\subsection{Singular-value statistics}

For a matrix $\mathbf{A}$, its singular values $\lambda$ are defined as the square roots of the 
eigenvalues of $\mathbf{A}\mathbf{A}^\dagger$ or $\mathbf{A}^\dagger\mathbf{A}$, where, as usual, 
$\mathbf{A}^\dagger$ is the conjugate transpose of $\mathbf{A}$. For a Hermitian matrix, 
i.e.~$\mathbf{A}^\dagger=\mathbf{A}$, the singular values reduce to the absolute values of the 
eigenvalues of $\mathbf{A}$.

In Ref.~\cite{KXOS23}, with the focus on open quantum systems, the statistical properties of singular 
values for all the 38-fold symmetry classes of non-Hermitian random matrices was extensively investigated.
There, it was shown that singular-value statistics (SVS) can be used as an effective measure for 
chaos and nonintegrability in open quantum systems. Specifically, the SVS of small random matrices
was analytically derived for the ratio between nearest neighbor singular values and for the minimum 
singular value; both were shown to describe well the SVS of large random matrices.

Given that the adjacency matrices $\mathbf{A}$ of directed random graphs and networks are 
non-Hermitian, it is straightforward to think on the SVS as a tool to study the spectral properties of 
$\mathbf{A}$ by computing the real singular values of 
$\mathbf{A}\mathbf{A}^\dagger$ instead of working with the complex eigenvalues of $\mathbf{A}$.
Indeed, that is the purpose of this work: Here we study the SVS of the non-Hermitian adjacency matrices 
of directed random graphs.
Since for real matrices, as the ones we consider here, the conjugate transpose is just the transpose
$\mathbf{A}^\dagger=\mathbf{A}^{\tbox{T}}$,
then, in what follows, the SVS concerns the spectra of $\mathbf{A}\mathbf{A}^{\tbox{T}}$.

In the next Section we characterize the real spectra of the Hermitian matrix 
$\mathbf{A}\mathbf{A}^{\tbox{T}}$ and, as a reference, we also characterize the complex spectra 
of the non-Hermitian adjacency matrix $\mathbf{A}$ by computing, respectively, the average value 
of the ratio between nearest neighbor singular values $r_\mathbb{R}$ and the average value of 
the ratio between nearest- and next-to-nearest neighbor eigenvalues $r_\mathbb{C}$, which are 
defined as follows.

On the one hand, given the real ordered spectrum 
$\lambda_1>\lambda_2>\cdots>\lambda_{n-1}>\lambda_n$, the $k$-th ratio $r_\mathbb{R}^k$ 
reads as~\cite{OH07,ABGR13}
\begin{equation}
\label{rR}
r_\mathbb{R}^k = \frac{\min(\lambda_{k+1}- \lambda_k,\lambda_{k}- \lambda_{k-1})}{\max(\lambda_{k+1}- \lambda_k,\lambda_{k}- \lambda_{k-1})} \ .
\end{equation}
Here, $r_\mathbb{R}\in[0,1]$.
On the other hand, given the complex spectrum $\{ \lambda_k \}$ the $k$-th ratio $r_\mathbb{C}^k$ 
reads as~\cite{SRP20}
\begin{equation}
r_\mathbb{C}^k = \frac{\left| \lambda^{NN}_k - \lambda_k \right|}{\left| \lambda^{NNN}_k - \lambda_k \right|} \ ,
\label{rC}
\end{equation}
where $\lambda^{NN}_k$ and $\lambda^{NNN}_k$ are, respectively, the nearest and the next-to-nearest 
neighbors of $\lambda_k$ in $\mathbb{C}$. Note that, as well as $r_\mathbb{R}$, $r_\mathbb{C}\in[0,1]$. 
Moreover, note that $r_\mathbb{C}$ can also be computed for real spectra.

It is relevant to mention that the singular values of certain adjacency matrices of specific deterministic 
graphs have already been studied in Ref.~\cite{P06}.
Also, there is an analytical study of the distribution of the minimum singular value of the randomly weighted 
adjacency matrices of directed Erd\"os-R\'enyi graphs in the regime of large average degree 
$\langle k \rangle \gg 1$~\cite{CL17}.
However, we are not aware of any statistical (numerical) study of the the singular values of random graphs.

\subsection{Models of directed random graphs}

We consider two models of directed random graphs $G$:
the directed Erd\"os-R\'enyi graph (dERG) model and the directed random regular graph (dRRG) model.
In the dERG model, $G(n,p)$ has $n$ vertices and each directed edge appears independently 
with probability $p \in (0,1]$.
While the graphs of the dRRG model, $G(n,\rho)$, consist of $n$ vertices uniformly and independently 
distributed on the unit square where two vertices are connected by a directed edge if their euclidean 
distance is smaller than the connection radius $\rho \in (0,\sqrt{2}]$.

\subsection{The randomly-weighted adjacency matrix}

Once a random directed graph is constructed, $G(n,p)$ or $G(n,\rho)$, its binary adjacency matrix is 
weighted with random variables (including self loops) as follows:
\begin{equation}
[\mathbf{A}]_{uv}=\left\{
\begin{array}{ll}
\epsilon_{uu} & \mbox{if $u=v$}, \\
\epsilon_{uv} & \mbox{if $u\rightarrow v$}, \\
0 & \mbox{otherwise}.
\end{array}
\right.
\label{A}
\end{equation}
Here, we choose $\epsilon_{uv}$ as statistically-independent random variables drawn from a 
normal distribution with zero mean and variance one, $\epsilon_{uv}\sim\mathcal {N}(0, 1)$. 
Evidently, since $G$ is directed, $\epsilon_{uv}\ne \epsilon_{vu}$; thus, $\mathbf{A}$ is non-Hermitian.
Note that $\mathbf{A}$ is a member of a {\it diluted} real Ginibre ensemble (RGE)~\cite{G65};
the RGE consists of random $n\times n$ matrices formed from independent and identically distributed 
standard Gaussian entries. Then, for a complete graph, when $p=1$ in the dERG model or 
$\rho=\sqrt{2}$ in the dRRG model, $\mathbf{A}$ becomes a member of the RGE. 
Notice, in addition, that when $p=0$ or $\rho=0$, i.e.~for isolated vertices, $\mathbf{A}$ becomes 
a member of the Poisson Ensemble (PE)~\cite{M04}; that is,~$\mathbf{A}$ is a diagonal real random 
matrix. Thus, a transition from the PE to the RGE is expected when increasing $p$ from zero to 
one in the dERG model and when increasing $\rho$ from zero to $\sqrt{2}$ in the dRRG model.
We note that some spectral properties of the RGE were reported in Ref.~\cite{FN07}. 
Also, diluted RGEs were already considered in Refs.~\cite{PRRCM20} and~\cite{PM23} as models 
of randomly weighted adjacency matrices of dERGs and dRRGs, respectively.

\section{Numerical results}

In what follows we use exact numerical diagonalization to obtain the eigenvalues $\lambda_k$ 
($k =1,\ldots,n$) of ensembles of adjacency matrices $\mathbf{A}$ and ensembles of the corresponding 
matrix products $\mathbf{A}\mathbf{A}^{\tbox{T}}$ for both $G(n,p)$ and $G(n,\rho)$.

\subsection{Ratio between nearest neighbor singular values}

\begin{figure}[t]
\centering
\includegraphics[width=0.95\columnwidth]{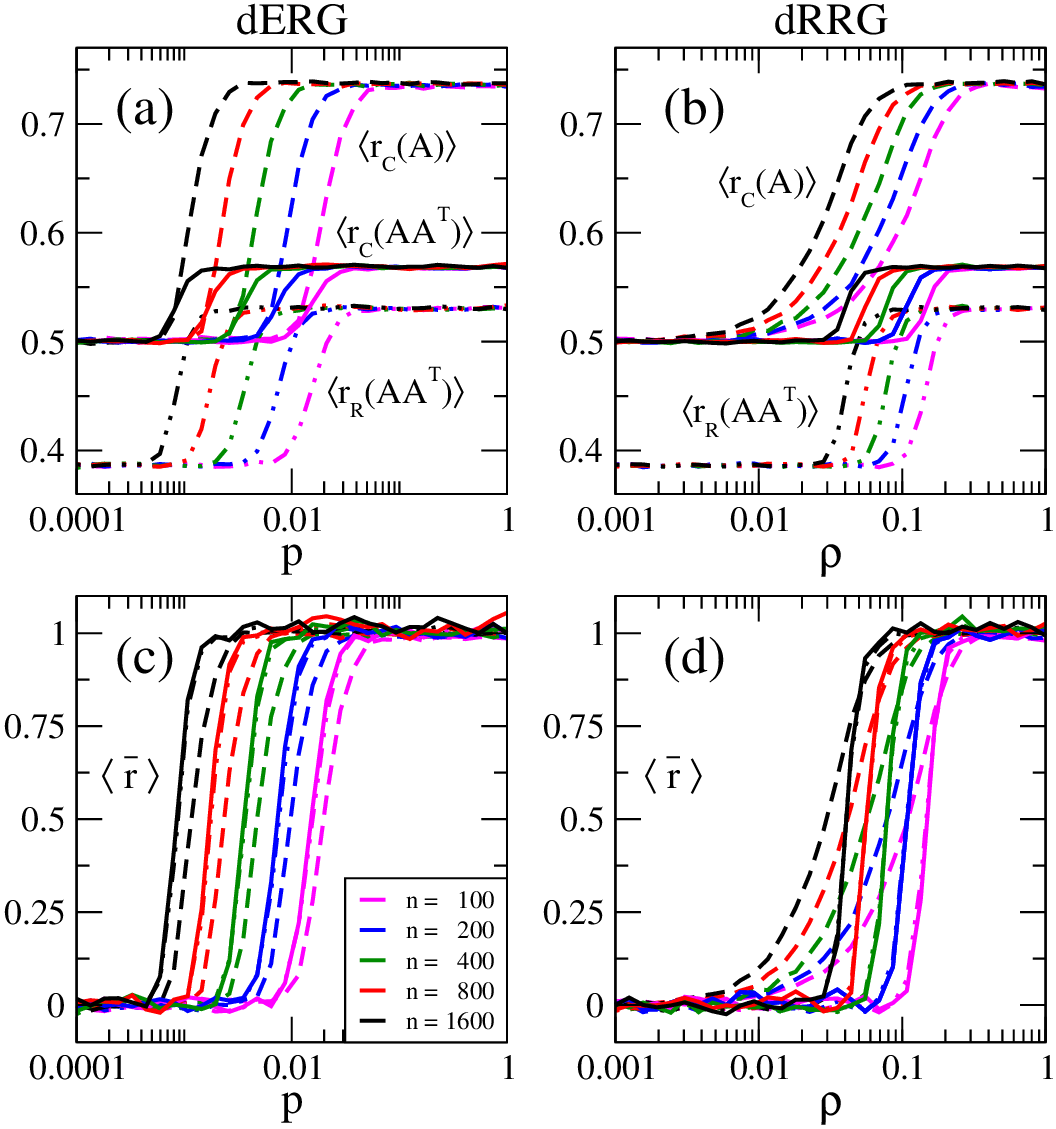}
\caption{Average ratios 
$\left\langle r_\mathbb{R}(\mathbf{A}\mathbf{A}^{\tbox{T}}) \right\rangle$ (dot-dashed lines), 
$\left\langle r_\mathbb{C}(\mathbf{A}\mathbf{A}^{\tbox{T}}) \right\rangle$ (full lines) 
and $\left\langle r_\mathbb{C}(\mathbf{A}) \right\rangle$ (dashed lines) of 
(a) directed Erd\"os-R\'enyi graphs (of size $n$) as a function of the probability $p$ and
(b) directed random regular graphs (of size $n$) as a function of the connection radius $\rho$.
Normalized ratios $\left\langle \overline{r}_\mathbb{R}(\mathbf{A}\mathbf{A}^{\tbox{T}}) \right\rangle$, 
$\left\langle \overline{r}_\mathbb{C}(\mathbf{A}\mathbf{A}^{\tbox{T}}) \right\rangle$ and
$\left\langle \overline{r}_\mathbb{C}(\mathbf{A}) \right\rangle$
[see Eqs.~(\ref{rRAA}), (\ref{rCAA}) and~(\ref{rCA}), respectively] for
(c) directed Erd\"os-R\'enyi graphs [same curves of panel (a)] and
(d) directed random regular graphs [same curves of panel (c)].
All averages are computed over $10^6$ ratios.}
\label{Fig01}
\end{figure}

In Fig.~\ref{Fig01}(a) we present the average ratio between nearest neighbor singular values 
$\left\langle r_\mathbb{R}(\mathbf{A}\mathbf{A}^{\tbox{T}}) \right\rangle$ (see dot-dashed lines)
of dERGs as a function of the probability $p$.
In Fig.~\ref{Fig01}(b) we also show 
$\left\langle r_\mathbb{R}(\mathbf{A}\mathbf{A}^{\tbox{T}}) \right\rangle$ (also as dot-dashed lines)
but of dRRGs as a function of the connection radius $\rho$.
For both graph models we used graphs of five sizes: $n=100$, 200, 400, 800, and 1600.
Indeed, we have verified that for these sizes the graph models are already in the large $n$
limit; see Appendix~\ref{appex} for a small-graph size analysis. 
For comparison purposes, in Figs.~\ref{Fig01}(a) and~\ref{Fig01}(b) we include 
$\left\langle r_\mathbb{C}(\mathbf{A}\mathbf{A}^{\tbox{T}}) \right\rangle$ (full lines)
and, as a reference, we also plot $\left\langle r_\mathbb{C}(\mathbf{A}) \right\rangle$ 
(dashed lines); i.e.~the ratio between nearest and next-to-nearest complex eigenvalues 
of the non-Hermitian adjacency matrix $\mathbf{A}$.
In fact, $\left\langle r_\mathbb{C}(\mathbf{A}) \right\rangle$ for dERGs and dRRGs has already 
been reported in Refs.~\cite{PRRCM20} and~\cite{PM23}, respectively.

\setlength{\tabcolsep}{6pt}
\begin{table}[b!]
\caption{Reference average values of the ratios $r_\mathbb{R}$ and $r_\mathbb{C}$ for the random 
adjacency matrices used in this work. To compute the averages, the spectra of $10^3$ adjacency 
matrices of size $n=1000$ were used.}
\label{T1}
\begin{tabular}{ c | c | c | c | c  }  
\hline
 & PE & (PE)(PE)$^{\tbox{T}}$ & RGE & (RGE)(RGE)$^{\tbox{T}}$  \\
\hline
$\left\langle r_\mathbb{R} \right\rangle$ 
& -- & 0.386 & -- & 0.531  \\
$\left\langle r_\mathbb{C} \right\rangle$  
& 0.500 & 0.500 & 0.737 & 0.569 \\
\hline
\end{tabular}
\end{table}
\begin{figure}[t]
\centering
\includegraphics[width=0.95\columnwidth]{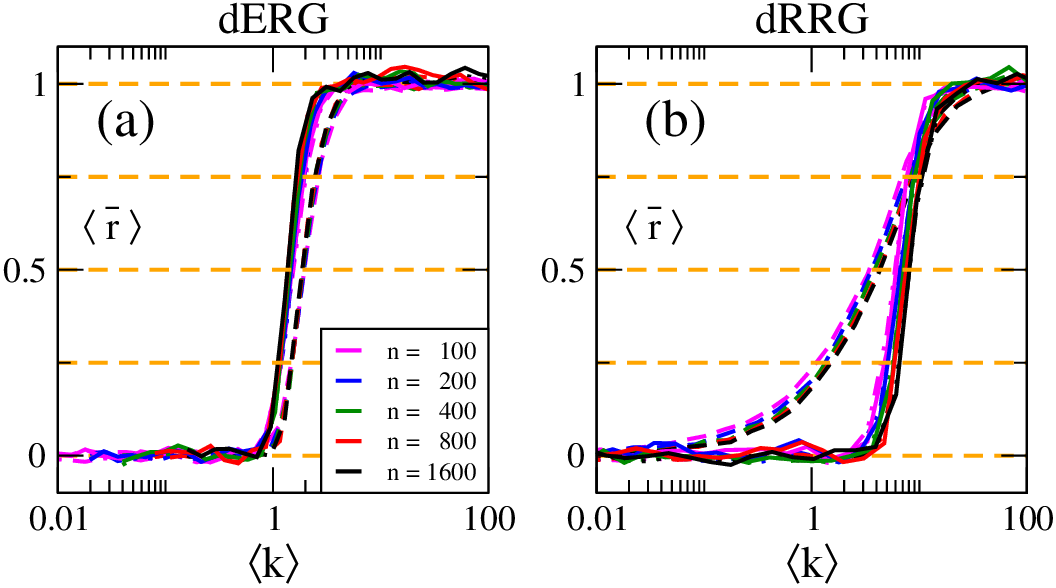}
\caption{
Normalized ratios $\left\langle \overline{r}_\mathbb{R}(\mathbf{A}\mathbf{A}^{\tbox{T}}) \right\rangle$, 
$\left\langle \overline{r}_\mathbb{C}(\mathbf{A}\mathbf{A}^{\tbox{T}}) \right\rangle$ and
$\left\langle \overline{r}_\mathbb{C}(\mathbf{A}) \right\rangle$ as a function of the average degree
$\left\langle k \right\rangle$ for
(a) directed Erd\"os-R\'enyi graphs [same curves of Fig.~\ref{Fig01}(c)] and
(b) directed random regular graphs [same curves of Fig.~\ref{Fig01}(d)].
Horizontal dashed lines in (a,b) mark the values of $\left\langle \overline{r} \right\rangle$ 
used to compute the PDFs reported in Fig.~\ref{Fig03}.}
\label{Fig06}
\end{figure}

To ease the comparison of the average ratios we conveniently normalize them as
\begin{equation}
    \langle \overline{r}_{\mathbb{R}}(\mathbf{A}\mathbf{A}^{\tbox{T}})\rangle \equiv \frac{\langle r_{\mathbb{R}}(\mathbf{A}\mathbf{A}^{\tbox{T}})\rangle -  \langle r_{\mathbb{R}}\rangle_{(\tbox{PE})(\tbox{PE})^{\tbox{T}}}}{ \langle r_{\mathbb{R}}\rangle_{(\tbox{RGE})(\tbox{RGE})^{\tbox{T}}}- \langle r_{\mathbb{R}}\rangle_{(\tbox{PE})(\tbox{PE})^{\tbox{T}}}},
    \label{rRAA}
\end{equation}
\begin{equation}
    \langle \overline{r}_{\mathbb{C}}(\mathbf{A}\mathbf{A}^{\tbox{T}})\rangle \equiv \frac{\langle r_{\mathbb{C}}(\mathbf{A}\mathbf{A}^{\tbox{T}})\rangle -  \langle r_{\mathbb{C}}\rangle_{(\tbox{PE})(\tbox{PE})^{\tbox{T}}}}{ \langle r_{\mathbb{C}}\rangle_{(\tbox{RGE})(\tbox{RGE})^{\tbox{T}}}- \langle r_{\mathbb{C}}\rangle_{(\tbox{PE})(\tbox{PE})^{\tbox{T}}}}
    \label{rCAA}
\end{equation}
and
\begin{equation}
    \langle \overline{r}_{\mathbb{C}}(\mathbf{A})\rangle \equiv \frac{\langle r_{\mathbb{C}}(\mathbf{A})\rangle -  \langle r_{\mathbb{C}}\rangle_{\tbox{PE}}}{ \langle r_{\mathbb{C}}\rangle_{\tbox{RGE}}- \langle r_{\mathbb{C}}\rangle_{\tbox{PE}}};
    \label{rCA}
\end{equation}
so they all take values between zero and one.
The reference values used in Eqs.~(\ref{rRAA}-\ref{rCA}), 
corresponding to the PE and the RGE, are reported in Table~\ref{T1}.
Then, in Figs.~\ref{Fig01}(c,d) we plot the normalized ratios 
$\left\langle \overline{r}_\mathbb{R}(\mathbf{A}\mathbf{A}^{\tbox{T}}) \right\rangle$,
$\left\langle \overline{r}_\mathbb{C}(\mathbf{A}\mathbf{A}^{\tbox{T}}) \right\rangle$ and 
$\left\langle \overline{r}_\mathbb{C}(\mathbf{A}) \right\rangle$
for both dERGs and dRRGs, respectively.

From Fig.~\ref{Fig01} we can conclude that all three ratios can
effectively characterize the transition between the regime of mostly isolated vertices, where
$\left\langle \overline{r}_\mathbb{R}(\mathbf{A}\mathbf{A}^{\tbox{T}}) \right\rangle 
\approx \left\langle \overline{r}_\mathbb{C}(\mathbf{A}\mathbf{A}^{\tbox{T}}) \right\rangle \approx 
\left\langle \overline{r}_\mathbb{C}(\mathbf{A}) \right\rangle \approx 0$,
and the regime of mostly connected graphs, where
$\left\langle \overline{r}_\mathbb{R}(\mathbf{A}\mathbf{A}^{\tbox{T}}) \right\rangle 
\approx \left\langle \overline{r}_\mathbb{C}(\mathbf{A}\mathbf{A}^{\tbox{T}}) \right\rangle \approx 
\left\langle \overline{r}_\mathbb{C}(\mathbf{A}) \right\rangle \approx 1$.
Moreover, notice that
\begin{equation}
\langle \overline{r}_{\mathbb{C}}(\mathbf{A}\mathbf{A}^{\tbox{T}})\rangle \approx
\langle \overline{r}_{\mathbb{R}}(\mathbf{A}\mathbf{A}^{\tbox{T}})\rangle
\label{rCrA}
\end{equation}
for both graph models; i.e.~the dot-dashed curves and the full curves in Figs.~\ref{Fig01}(c,d) fall one 
on top the other (for a give graph size $n$ and a given graph model). 
This means that $\left\langle r_\mathbb{R}(\mathbf{A}\mathbf{A}^{\tbox{T}}) \right\rangle$ and 
$\left\langle r_\mathbb{C}(\mathbf{A}\mathbf{A}^{\tbox{T}}) \right\rangle$, once normalized, provide 
the same information; so they can be used indistinguishably on real spectra.

\begin{figure*}[!t]
\includegraphics[width=0.8\textwidth]{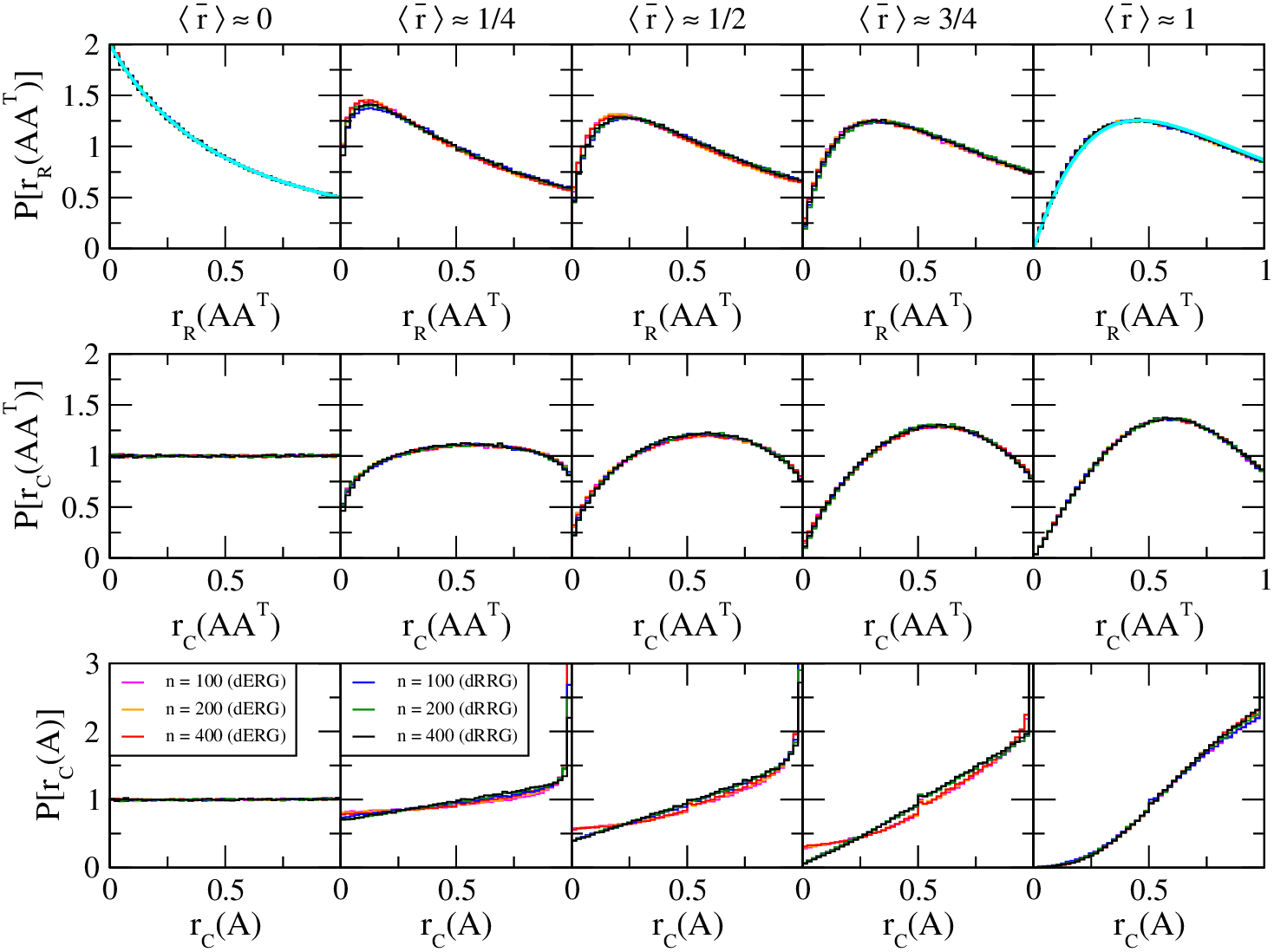}
\caption{Probability density function of the ratios 
$r_\mathbb{R}(\mathbf{A}\mathbf{A}^{\tbox{T}})$ (upper panels), 
$r_\mathbb{C}(\mathbf{A}\mathbf{A}^{\tbox{T}})$ (middle panels) and
$r_\mathbb{C}(\mathbf{A})$ (lower panels)
of directed Erd\"os-R\'enyi graphs and directed random regular graphs of size $n$.
Each histogram was constructed from the ratios of $10^6$ random graphs.
The normalized rations $\langle \overline{r}\rangle$ are fixed in each column.
Full cyan lines in upper-left and upper-right panels are 
Eqs.~(\ref{PrPE}) and (\ref{PrGOE}), respectively.}
\label{Fig03}
\end{figure*}

It is also instructive to plot the ratios as a function of the average degree
$\left\langle k \right\rangle$, see Fig.~\ref{Fig06}; so we can better contrast both graph models.
Here, we can clearly see that the transition from almost isolated vertices to almost complete
graphs is quite sharp when characterized by
$\left\langle r_\mathbb{R}(\mathbf{A}\mathbf{A}^{\tbox{T}}) \right\rangle$ and 
$\left\langle r_\mathbb{C}(\mathbf{A}\mathbf{A}^{\tbox{T}}) \right\rangle$ for both graph models.
However, when using $\left\langle r_\mathbb{C}(\mathbf{A}) \right\rangle$, the transition is 
much wider for dRRGs than for dERGs.
This may indicate that $\left\langle r_\mathbb{C}(\mathbf{A}) \right\rangle$ may be a better
tool to distinguish different graph models than 
$\left\langle r_\mathbb{R}(\mathbf{A}\mathbf{A}^{\tbox{T}}) \right\rangle$ or 
$\left\langle r_\mathbb{C}(\mathbf{A}\mathbf{A}^{\tbox{T}}) \right\rangle$.
Anyway, it could be possible to distinguish between the graph models we are exploring here
by using $\left\langle r_\mathbb{R}(\mathbf{A}\mathbf{A}^{\tbox{T}}) \right\rangle$ or 
$\left\langle r_\mathbb{C}(\mathbf{A}\mathbf{A}^{\tbox{T}}) \right\rangle$: Note that the
transition for dERGs starts at a smaller value of $\left\langle k \right\rangle$ 
($\left\langle k \right\rangle\approx 1$) as compared to dRRGs ($\left\langle k \right\rangle\approx 4$).

For completeness, we now inspect the probability density functions (PDFs) of the ratios 
$r_\mathbb{R}(\mathbf{A}\mathbf{A}^{\tbox{T}})$, 
$r_\mathbb{C}(\mathbf{A}\mathbf{A}^{\tbox{T}})$ and
$r_\mathbb{C}(\mathbf{A})$.
Moreover, for a meaningful comparison, we compute the PDFs at fixed 
values of the normalized ratios. Specifically, we choose five values of $\langle \overline{r}\rangle$, 
as indicated by the horizontal dashed lines in Figs.~\ref{Fig06}(a,b):
$\langle \overline{r}\rangle\approx 0$, 1/4, 1/2, 3/4, and 1.
We note that there are many possible values of $p$ [$\rho$] that produce 
$\langle \overline{r}\rangle\approx 0$ and 1 for dERGs [dRRGs]; here and below we set 
 $p=0.0001$ [$\rho=0.001$] and $p=0.5$ [$\rho=1$] to get 
$\langle \overline{r}\rangle\approx 0$ and 1, respectively.

Then, in the upper panels of Fig.~\ref{Fig03} we report the probability density function of the ratio 
$r_\mathbb{R}(\mathbf{A}\mathbf{A}^{\tbox{T}})$. Each panel corresponds to a fixed average ratio
$\langle \overline{r}\rangle$.
Note that each panel in Fig.~\ref{Fig03} contains six histograms: three of dERGs 
(in light colors) and three of dRRGs (in dark colors) of different sizes.
As expected, once the average ratio $\langle \overline{r}\rangle$ is fixed, 
$P[r_\mathbb{R}(\mathbf{A}\mathbf{A}^{\tbox{T}})]$
does not depend on the graph size, and in this case, neither on the graph model.
Also, as expected~\cite{KXOS23}, when $\langle \overline{r}\rangle\approx 0$, 
$P[r_\mathbb{R}(\mathbf{A}\mathbf{A}^{\tbox{T}})]$ is well reproduced by the RMT prediction for the
PE~\cite{ABGR13} 
\begin{equation}
P_{\tbox{PE}}(r_\mathbb{R}) = \frac{2}{(1+r_\mathbb{R})^2} \ ,
\label{PrPE}
\end{equation}
see the cyan curve in the upper-left panel;
while for $\langle \overline{r}\rangle\approx 1$, $P[r_\mathbb{R}(\mathbf{A}\mathbf{A}^{\tbox{T}})]$ 
corresponds to the RMT prediction for the GOE~\cite{ABGR13}
\begin{equation}
P_{\tbox{GOE}}(r_\mathbb{R}) = 
\frac{27}{4} \frac{r_\mathbb{R}+r_\mathbb{R}^2}{(1+r_\mathbb{R}+r_\mathbb{R}^2)^{5/2}} \ ,
\label{PrGOE}
\end{equation}
see the cyan curve in the upper-right panel.

In Fig.~\ref{Fig03} we also plot $P[r_\mathbb{C}(\mathbf{A}\mathbf{A}^{\tbox{T}})]$ (middle panels) 
and $P[r_\mathbb{C}(\mathbf{A})]$ (lower panels). It is interesting to note that, as well as 
for $P[r_\mathbb{R}(\mathbf{A}\mathbf{A}^{\tbox{T}})]$, once the average ratio is fixed, 
$P[r_\mathbb{C}(\mathbf{A}\mathbf{A}^{\tbox{T}})]$ can not distinguish between different graph
models; i.e.~all histograms in the upper and middle panels of Fig.~\ref{Fig03} fall one on top of the other.
However, $P[r_\mathbb{C}(\mathbf{A})]$ can indeed distinguish between dERGs and dRRGs: 
Notice that the histograms corresponding to different graph models
follow slightly different shapes; see the difference between light-color (dERGs) and dark-color (dRRGs) 
histograms in the lower panels of Fig.~\ref{Fig03}), specifically when $0<\langle \overline{r}\rangle<1$.
This is in accordance with the observations made in Fig.~\ref{Fig06}:
While $\left\langle r_\mathbb{R}(\mathbf{A}\mathbf{A}^{\tbox{T}}) \right\rangle$ and 
$\left\langle r_\mathbb{C}(\mathbf{A}\mathbf{A}^{\tbox{T}}) \right\rangle$ are not able to easily
distinguish between the two graph models, $\left\langle r_\mathbb{C}(\mathbf{A}) \right\rangle$ can.

\subsection{Minimum singular value $\lambda_{\tbox{min}}$}

\begin{figure}[ht]
\centering
\includegraphics[width=0.95\columnwidth]{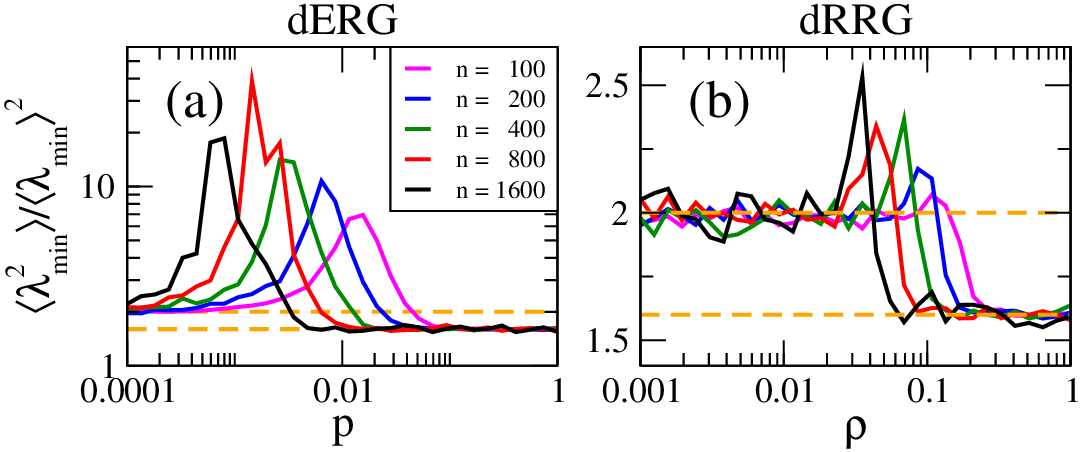}
\caption{(a) [(b)] 
$\left\langle \lambda_{\tbox{min}}^2 \right\rangle/\left\langle \lambda_{\tbox{min}} \right\rangle^2$ 
of directed Erd\"os-R\'enyi graphs of size $n$ as a function of the probability $p$
[of directed random regular graphs of size $n$ as a function of the connection radius $\rho$].
Orange dashed lines correspond to $\left\langle \lambda_{\tbox{min}}^2 \right\rangle_{\tbox{PE}}/\left\langle \lambda_{\tbox{min}} \right\rangle_{\tbox{PE}}^2=2$~\cite{KXOS23} and 
$\left\langle \lambda_{\tbox{min}}^2 \right\rangle_{\tbox{RGE}}/\left\langle \lambda_{\tbox{min}} \right\rangle_{\tbox{RGE}}^2\approx 1.6$.
The averages are computed over $10^6/n$ minimum singular values.}
\label{Fig04}
\end{figure}
\begin{figure*}[!t]
\includegraphics[width=0.8\textwidth]{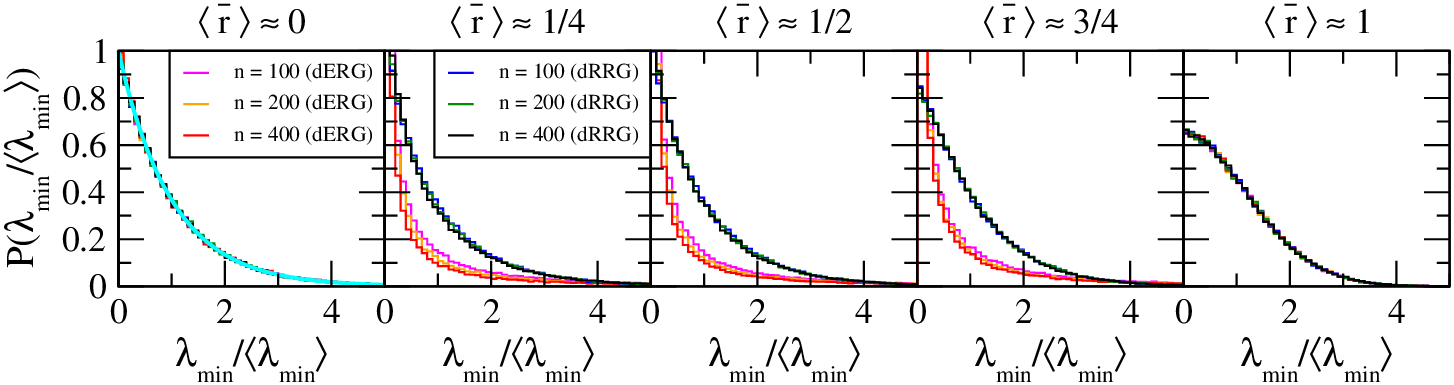}
\caption{
Probability density function of the minimum singular value $\lambda_{\tbox{min}}$
normalized to $\left\langle \lambda_{\tbox{min}} \right\rangle$
of directed Erd\"os-R\'enyi graphs and directed random regular graphs of size $n$.
Each histogram was constructed from the ratios of $10^5$ random graphs.
The normalized rations $\langle \overline{r}\rangle$ are fixed in each panel.
Full cyan line in left panel is 
$P_{\tbox{PE}}(\lambda_{\tbox{min}}/\left\langle \lambda_{\tbox{min}} \right\rangle)=
\exp[-\lambda_{\tbox{min}}/\left\langle \lambda_{\tbox{min}} \right\rangle]$~\cite{KXOS23}.}
\label{Fig05}
\end{figure*}

Now we explore the statistics of the minimum singular value $\lambda_{\tbox{min}}$.
We start by plotting, in Figs.~\ref{Fig04}(a) and~\ref{Fig04}(b), the ratio
$\left\langle \lambda_{\tbox{min}}^2 \right\rangle/\left\langle \lambda_{\tbox{min}} \right\rangle^2$ 
for dERGs (as a function of the probability $p$) and for dRRGs (as a function of the connection radius 
$\rho$), respectively. 
From Fig.~\ref{Fig04} we observe, by increasing the connectivity in both graph models, the transition of  
$\left\langle \lambda_{\tbox{min}}^2 \right\rangle/\left\langle \lambda_{\tbox{min}} \right\rangle^2$ 
from the PE value 
$\left\langle \lambda_{\tbox{min}}^2 \right\rangle_{\tbox{PE}}/\left\langle \lambda_{\tbox{min}} \right\rangle_{\tbox{PE}}^2=2$~\cite{KXOS23} to the RGE value, that we report as
$\left\langle \lambda_{\tbox{min}}^2 \right\rangle_{\tbox{RGE}}/\left\langle \lambda_{\tbox{min}} \right\rangle_{\tbox{RGE}}^2\approx 1.6$. We note that our value for $\left\langle \lambda_{\tbox{min}}^2 \right\rangle_{\tbox{RGE}}/\left\langle \lambda_{\tbox{min}} \right\rangle_{\tbox{RGE}}^2$ is larger than that analytically computed in Ref.~\cite{KXOS23}
for $2\times 2$ matrices; this difference is due to the graph sizes we consider here, that could be considered
as large sizes.
However, as it is clear from this figure, the PE--to--RGE transition is not smooth and for increasing connectivity
the ratio $\left\langle \lambda_{\tbox{min}}^2 \right\rangle/\left\langle \lambda_{\tbox{min}} \right\rangle^2$
first increases, reaches a maximum and then decreases approaching the RGE value.
What is even more remarkable is that for dERGs the maximal values that 
$\left\langle \lambda_{\tbox{min}}^2 \right\rangle/\left\langle \lambda_{\tbox{min}} \right\rangle^2$
can reach are about one order of magnitude larger that those for dRRGs. So, the ratio
$\left\langle \lambda_{\tbox{min}}^2 \right\rangle/\left\langle \lambda_{\tbox{min}} \right\rangle^2$
can indeed distinguish between both graph models.

Finally, in Fig.~\ref{Fig05} we report the probability density function of the minimum singular value 
$\lambda_{\tbox{min}}$ normalized to $\left\langle \lambda_{\tbox{min}} \right\rangle$, 
$P(\lambda_{\tbox{min}}/\left\langle \lambda_{\tbox{min}} \right\rangle)$, of dERGs and dRRGs.
As in Fig.~\ref{Fig03}, here each panel of Fig.~\ref{Fig05} contains six histograms: three of dERGs 
(in light colors) and three of dRRGs (in dark colors) of different sizes. Also each panel corresponds 
to a fixed average ratio $\langle \overline{r}\rangle$ (as indicated on top of the panels).
As expected, we observe the PE--to--RGE transition of 
$P(\lambda_{\tbox{min}}/\left\langle \lambda_{\tbox{min}} \right\rangle)$ for increasing  
connectivity in both graph models, here parametrized by $\langle \overline{r}\rangle$; see the
limiting PDFs in the left and right panels of Fig.~\ref{Fig05}.
However, for intermediate values of $\langle \overline{r}\rangle$, see the central panels of Fig.~\ref{Fig05},
$P(\lambda_{\tbox{min}}/\left\langle \lambda_{\tbox{min}} \right\rangle)$ is clearly different
for different graph models: Notice that the histograms for dERGs (in light colors) do not coincide
with those for dRRGs (in dark colors). Thus, as well as the ratio 
$\left\langle \lambda_{\tbox{min}}^2 \right\rangle/\left\langle \lambda_{\tbox{min}} \right\rangle^2$,
$P(\lambda_{\tbox{min}}/\left\langle \lambda_{\tbox{min}} \right\rangle)$ can clearly distinguish between 
both graph models.

\section{Discussion and conclusions}

We have performed a numerical study of the singular-value statistics (SVS) of the non-Hermitian, randomly 
weighted, adjacency matrices $\mathbf{A}$ of two models of random graphs: directed Erd\"os-R\'enyi graphs 
(dERGs) and directed random regular graphs (dRRGs).

We have computed the ratios $r$ between nearest neighbor singular values and the minimum singular 
values $\lambda_{\tbox{min}}$. We have shown that the average values of $r$ and $\lambda_{\tbox{min}}$ 
as well as the corresponding PDFs can effectively characterize the transition between mostly isolated vertices 
(Poisson Ensemble (PE) regime) to almost complete graphs (Real Ginibre Ensemble (RGE) regime);
see Figs.~\ref{Fig06} to~\ref{Fig05}.
However, in contrast to $\left\langle r \right\rangle$ and $P(r)$, both 
$\left\langle \lambda_{\tbox{min}} \right\rangle$ and $P(\lambda_{\tbox{min}})$ can clearly distinguish 
between the two graph models; see Figs.~\ref{Fig04} and~\ref{Fig05}.
This means that even though both graph models produce adjacency matrices $\mathbf{A}$ which are 
members of diluted RGEs which are structurally very similar, such that most average topological quantities
can not distinguish them (see e.g.~\cite{AMRS21}), the minimum singular values can.


In addition, there may be other RMT tools that could also be incorporated in the study and characterization 
of directed graphs and networks, such as the hard-edge statistics, recently discussed in Ref.~\cite{XSK24}.

\appendix 

\section{Small graphs}
\label{appex}

\begin{figure}[ht]
\centering
\includegraphics[width=0.95\columnwidth]{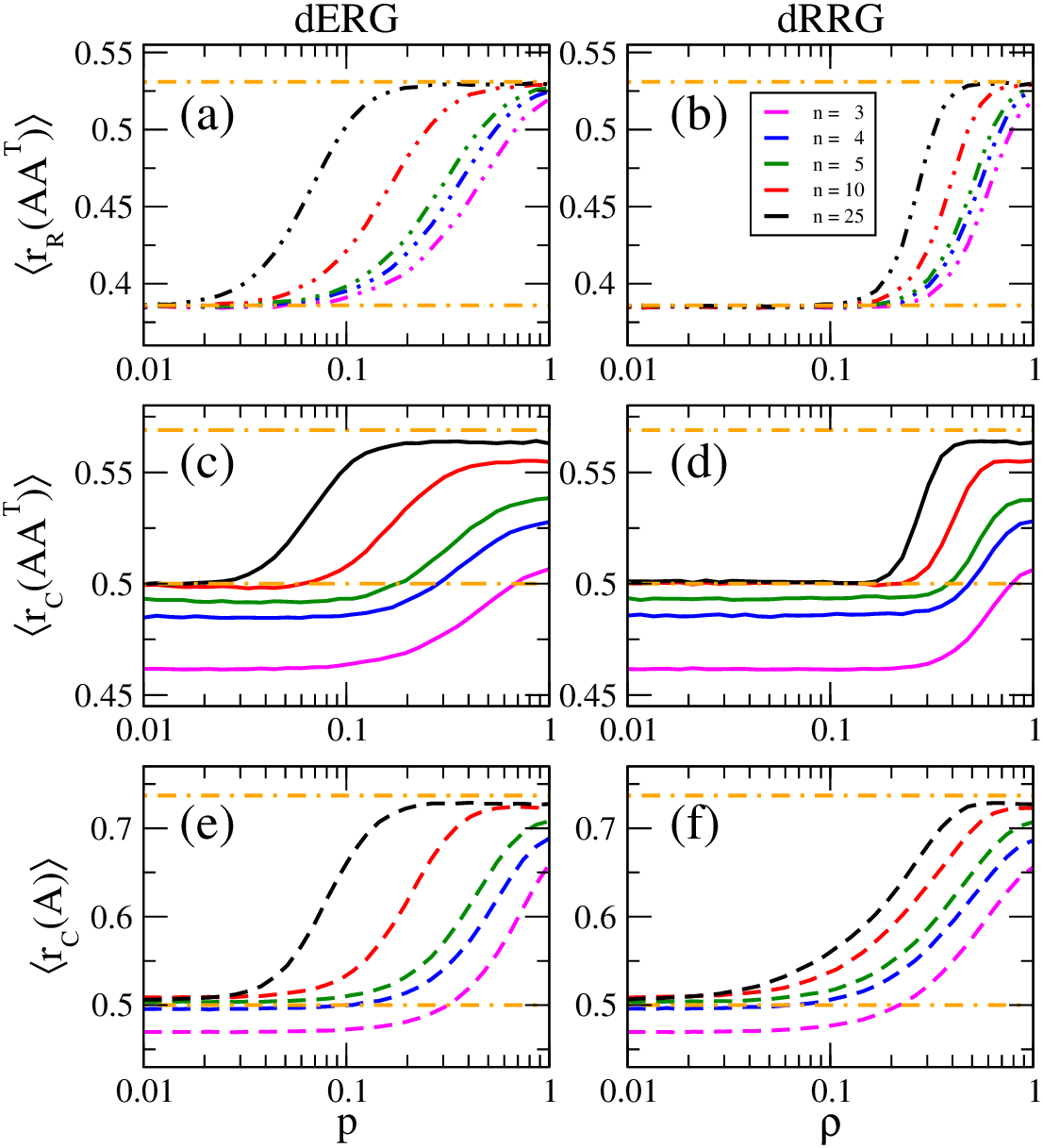}
\caption{Average ratios 
(a) $\left\langle r_\mathbb{R}(\mathbf{A}\mathbf{A}^{\tbox{T}}) \right\rangle$, 
(b) $\left\langle r_\mathbb{C}(\mathbf{A}\mathbf{A}^{\tbox{T}}) \right\rangle$ and
(c) $\left\langle r_\mathbb{C}(\mathbf{A}) \right\rangle$
of directed Erd\"os-R\'enyi graphs (of small size $n$) as a function of the probability $p$.
Average ratios 
(d) $\left\langle r_\mathbb{R}(\mathbf{A}\mathbf{A}^{\tbox{T}}) \right\rangle$,  
(e) $\left\langle r_\mathbb{C}(\mathbf{A}\mathbf{A}^{\tbox{T}}) \right\rangle$ and
(f) $\left\langle r_\mathbb{C}(\mathbf{A}) \right\rangle$
of directed random regular graphs (of small size $n$) as a function of the connection radius $\rho$.
All averages are computed over $10^6$ ratios.
Horizontal dashed lines are the expected values of the ratios for the PE (lower lines) and
the RGE (upper lines) for large graphs, as given in Table~\ref{T1}.}
\label{Fig02}
\end{figure}

In order to make sure that the results reported in the main text are already in the 
large graph-size limit, here we perform a study of small graphs.
Then, in Fig.~\ref{Fig02} we plot 
$\left\langle r_\mathbb{R}(\mathbf{A}\mathbf{A}^{\tbox{T}}) \right\rangle$,  
$\left\langle r_\mathbb{C}(\mathbf{A}\mathbf{A}^{\tbox{T}}) \right\rangle$ and
$\left\langle r_\mathbb{C}(\mathbf{A}) \right\rangle$ of 
dERGs (left panels) and dRRGs (right panels) of small size $n$.
In all panels we include, as horizontal dashed lines, the expected values of the ratios for the PE 
(lower lines) and the RGE (upper lines) for large graphs, as given in Table~\ref{T1}.

From Fig.~\ref{Fig02} we can clearly see that while 
$\left\langle r_\mathbb{R}(\mathbf{A}\mathbf{A}^{\tbox{T}}) \right\rangle$ reproduces both the
PE and the RGE values expected for large graphs already for graph sizes of the order of $n=10$,
see red curves in panels (a,b), for $\left\langle r_\mathbb{C}(\mathbf{A}\mathbf{A}^{\tbox{T}}) \right\rangle$ 
and $\left\langle r_\mathbb{C}(\mathbf{A}) \right\rangle$ larger graphs sizes are needed.
We concluded that to avoid small-graph size effects, for all ratios, we need to set $n\ge 50$ at least.
Therefore, for the calculations in the main text we consider $n\ge 100$.

\section*{Acknowledgements}
J.A.M.-B. thanks support from CONAHCyT-Fronteras (Grant No.~425854) 
and VIEP-BUAP (Grant No.~100405811-VIEP2024), Mexico.


\end{document}